\documentclass[prb,preprint,showpacs,preprintnumbers,amsmath,amssymb]{revtex4}
\usepackage{graphicx}
\usepackage{dcolumn}
\usepackage{color}
\usepackage[usenames,dvipsnames]{xcolor}

\begin{document}

\title{Extended Pattern Recognition Scheme for Self-learning Kinetic Monte Carlo (SLKMC-II) Simulations}

\email{abdelkader.kara@ucf.edu}
\author{Syed Islamuddin Shah}
\author{Giridhar Nandipati}
\email{giridhar.nandipati@ucf.edu	}
\author{Abdelkader Kara}
\email{islamuddin@knights.ucf.edu}
\author{Talat S. Rahman}
\email{talat.rahman@ucf.edu}
\affiliation{Department of Physics, University of Central Florida,  Orlando, FL  32816}
\date{\today}
\begin{abstract}
We report the development of a pattern-recognition scheme that takes into account both fcc and hcp adsorption sites in performing self-learning kinetic Monte Carlo (SLKMC-II) simulations on the fcc(111) surface. In this scheme, the local environment of every under-coordinated atom in an island is uniquely identified by grouping fcc sites, hcp sites and top-layer substrate atoms around it into hexagonal rings. As the simulation progresses, all possible processes including those like shearing, reptation and concerted gliding, which may involve fcc-fcc, hcp-hcp and fcc-hcp moves are automatically found, and their energetics calculated on the fly. In this article we present the results of applying this new pattern-recognition scheme to the self-diffusion of 9-atom islands ($M_{9}$) on M(111), where M = Cu, Ag or Ni.
\end{abstract}
\maketitle
\section{Introduction}
Kinetic Monte Carlo (KMC)\cite{bkl,gilmer,voter1,maksym,kristen, Blue} method is an extremely efficient method for carrying out dynamical simulations of a wide variety of stochastic and/or thermally-activated processes when these are known in advance. Accordingly, KMC simulations have been successfully used to model a variety of dynamical processes ranging from catalysis to thin-film growth. However in many cases it is difficult to know {\it a priori} all the relevant processes that may be important during simulation. To overcome this problem, on-the fly KMC\cite{akmc1, neb} methods were developed that allow the calculation of all possible processes at each KMC step. But the fact that on-the fly KMC methods do not store these calculated processes results in redundancies and repetitions in the calculations of energetics of the system dynamics. In order to avoid repeated calculation of energetics of the processes previously encountered, self-learning KMC (SLKMC) method\cite{slkmc1} was developed, which introduces a pattern-recognition scheme that allows the on-the-fly identification, storage and retrieval of information about processes based on the local neighborhood of each active (under-coordinated) atom speeding up simulations by several orders of magnitude. 

On the fcc(111) surface there are two types of threefold adsorption sites: the normal fcc site and the fault hcp site. The pattern recognition scheme used in previous studies\cite{slkmc1,slkmc2,pslkmc, iss_short} for fcc(111) surface was restricted to the occupation of fcc sites only and hence was unable to account for the processes that involve movement of atoms to or from hcp sites.\cite{repp,chang}. In this paper we present a pattern-recognition scheme that does allow occupation of hcp as well as fcc sites on the fcc(111) surface. We then illustrate its use by applying it to the self-diffusion of three different systems. 

The organization of the paper is as follows: In section~\ref{PRS} we present details of the pattern-recognition scheme and of its implementation for the fcc(111) surface. In section~\ref{results} we present results of the application of SLKMC-II to the self-diffusion of a 9-atom Cu, Ag, or Ni island on the corresponding fcc(111) surface; we also present examples of some processes that were found during our simulations. In section~\ref{SUMMARY AND PROSPECTS} we offer our conclusions.
\section{PATTERN RECOGNITION SCHEME}\label{PRS}
In order to accommodate hcp as well as fcc sites, SLKMC-II modifies both the pattern recognition scheme and the saddle-point search for finding processes and calculating their energetics.
A pattern-recognition scheme allows unique identification of the local neighborhood of an atom by assigning a unique combination of an index number or key, enabling SLKMC simulations to store and retrieve on the fly information about processes from a database that in traditional KMC simulations must be hardwired. 

As mentioned earlier, an adatom island on an fcc(111) surface may occupy either an fcc or an hcp site or a combination of fcc and hcp sites, but the pattern-recognition scheme used in SLKMC simulations up to now is limited to identifying adatom islands on fcc sites only. We have now improved the scheme to enable identification of adatom islands occupying hcp sites as well. This pattern-recognition scheme is simple, fast and capable of handling 2-D on-lattice fcc(111) systems. We call this SLKMC method with new pattern recognition scheme SLKMC-II.
\begin{figure}
\center{\includegraphics [width=8.5cm]{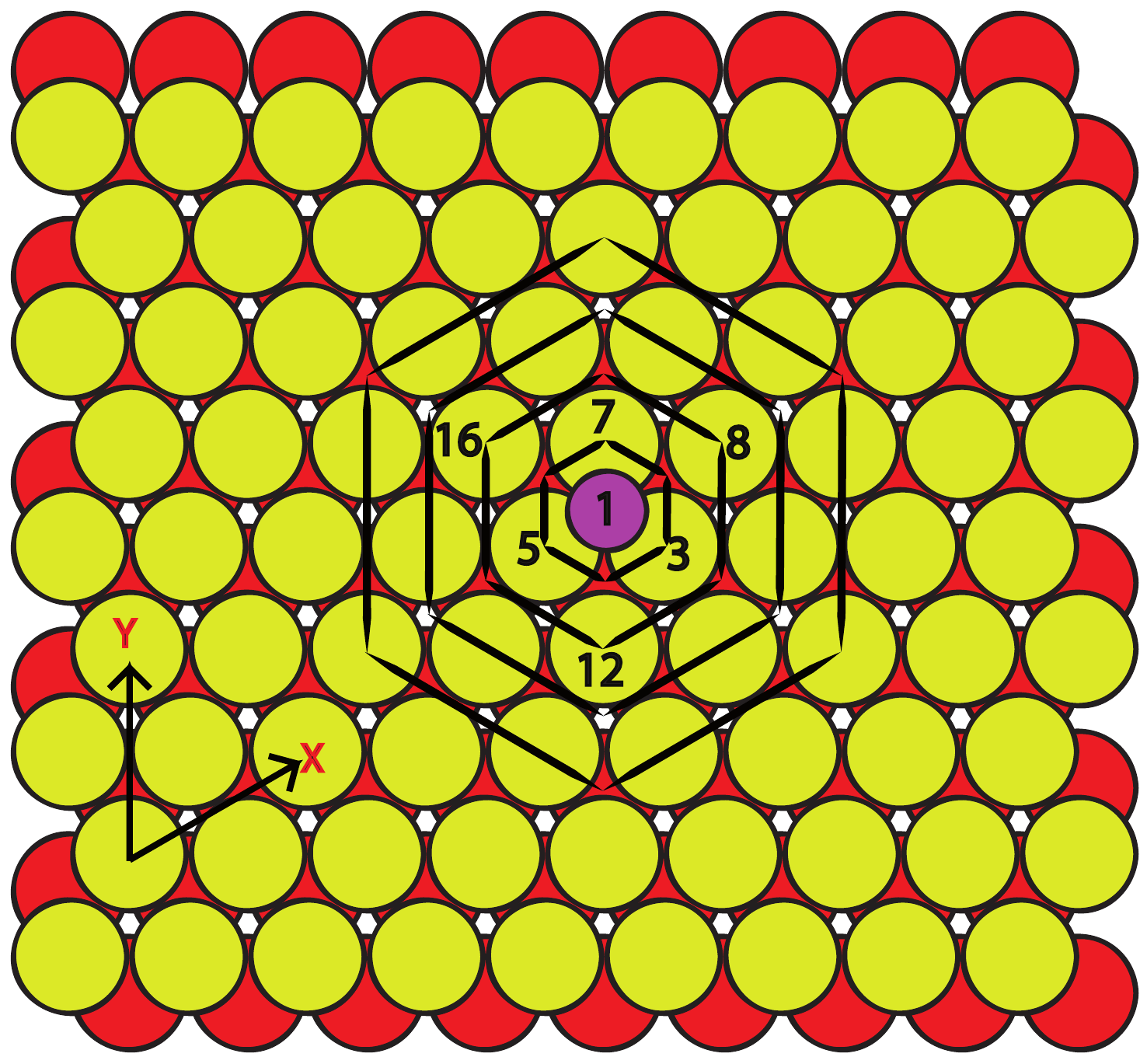} 
\caption{\label{fig1}{Grouping different sites into hexagonal rings on fcc(111) surface.}}}
\end{figure}
Like the previous fcc-only scheme, the pattern-recognition scheme used in SLKMC-II groups adsorption sites into hexagonal rings, generates a unique binary bit-pattern based on the occupancy of those sites, and stores it in a database along with the processes associated with that configuration. In order to uniquely identify whether an atom is on an fcc or hcp site, we include the top-layer substrate atoms (henceforth referred to as substrate atoms) in the pattern-recognition scheme. 

Fig.~\ref{fig1} shows the grouping of fcc, hcp and substrate atoms into the first 4 such hexagonal rings around a monomer (represented by the purple-colored circle) on an fcc site marked as 1. It can be seen that, except for the first ring, which is a combination of hcp sites and the substrate atoms, the rings are combinations of fcc sites, hcp sites and substrate atoms. We note that for a monomer on an hcp site, the first ring is a combination of fcc sites and substrate atoms.
\begin{figure}
\center{\includegraphics [width=8.5cm]{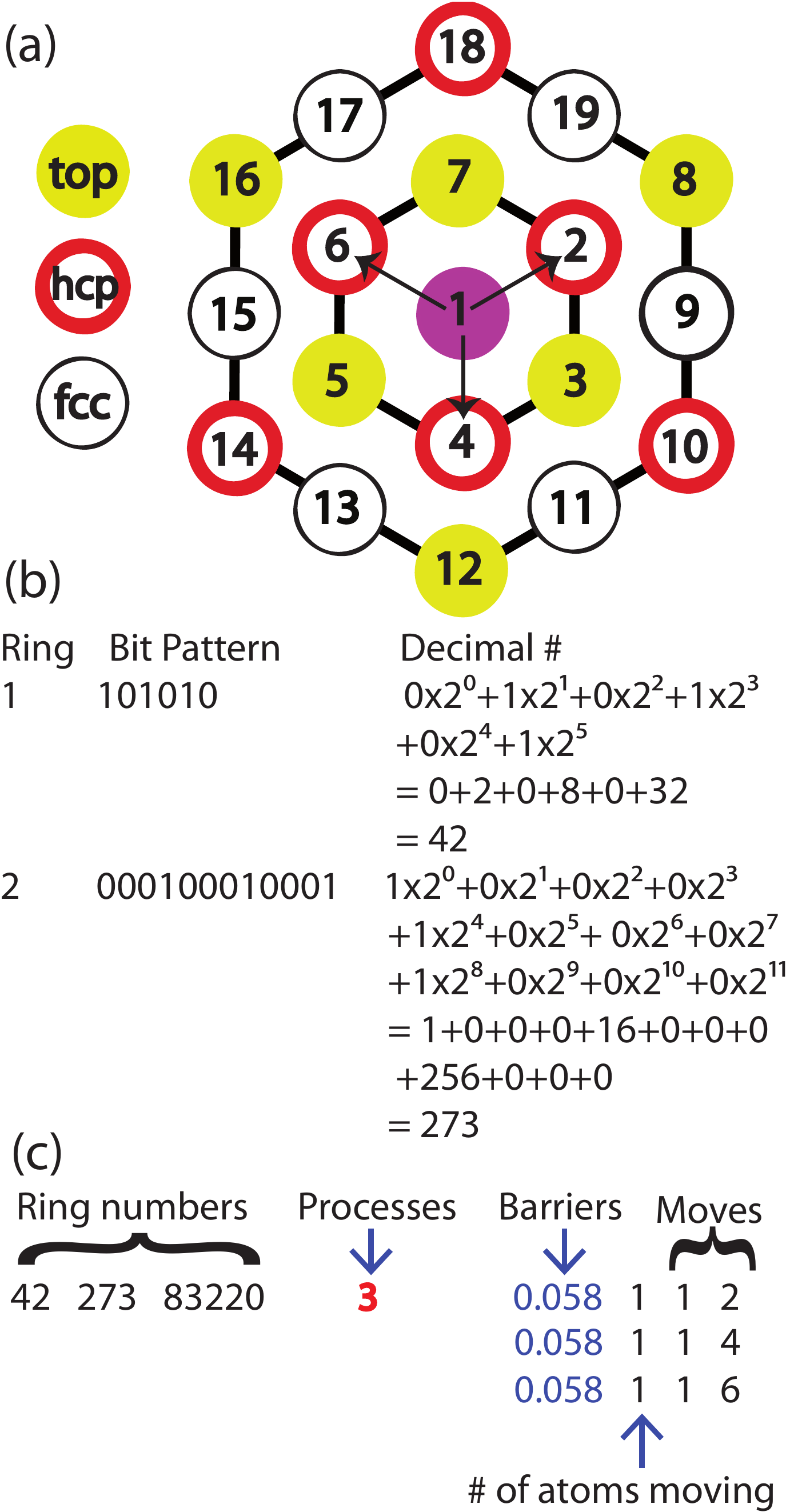} 
\caption{\label{fig2}{(a) Gold (light filled) circles represents substrate atoms while black and red (thick) circles represent fcc and hcp sites respectively. Purple (dark filled circle) represents an atom on fcc site while arrows show possible processes for such an atom. (b) Assignment of binary bit pattern and corresponding conversion into decimal number. (c) Format of the database for this configuration.}}}
\end{figure}

The X and Y arrows in Fig.~\ref{fig1} show the directions on the fcc(111) surface that we map onto a square lattice for our simulations. To represent the fcc(111) surface, we assign each substrate atom a height of 1, each unoccupied fcc or hcp site a height of $0$, and each occupied fcc or hcp site a height of 2. (Note that any site to the left of a substrate atom along the X-axis in Fig.~\ref{fig1} is an hcp site, while any site to the right of the substrate atom along that axis is an fcc site.) The set of rings in Fig.~\ref{fig1} are further elaborated in Fig.~\ref{fig2}(a) with numbering scheme that we have used. With the diffusing atom at an fcc site labeled as 1, we mark the hcp sites and substrate atoms surrounding it in ring 1 clockwise starting from the X-axis from 2 to 7, as shown in Fig.~\ref{fig2}(a). The sites in the second ring are similarly labeled from 8 to 19. The same procedure follows for the other rings. In the binary bit pattern (cf. Fig.~\ref{fig2}(b)), substrate atoms are always represented as 1, while fcc or hcp sites are represented as either 1 or 0 depending on whether the site is occupied or not. As in the numbering of the sites, the binary digits for a ring (cf. Fig.~\ref{fig2}(b)) begin at the X-axis (cf. Fig.~\ref{fig1}) and proceed clockwise. Resulting binary bit sequence is recorded (starting from right to left) as shown in Fig. ~\ref{fig2}(b). Ring numbers for a monomer at an fcc site in a three-ring system are as shown in Fig. ~\ref{fig2}(b $\&$ c). For convenience we have shown only two rings, although more rings are required to cover the neighborhood of larger islands. Following this method, ring numbers for every active atom in an island are generated on the basis of the occupancy of sites in its surrounding rings. For each configuration thus identified, a saddle-point search is initiated to find all possible processes and the activation energy for each. The result: configuration, its processes, and their activation energies are then stored in the database. For the case of monomer on an fcc site as shown in Fig.~\ref{fig2}(c), the configuration is represented by three rings as $42$, $273$, $83220$, it has three processes each having an activation barrier of 0.058 eV, for each process only one atom is involved and finally the actual move for process 1 is that atom at position 1 moves to position 2.

An atom is considered to be active if there is at least one vacant site in the second ring. As in the previous (fcc-only) pattern-recognition scheme, a process is identified as the motion of an active atom to a neighboring ring, accompanied by the motion of any other atom or atoms in the island. We note that in this scheme an atom can move more than one ring, namely when it moves from an fcc to a nearest fcc site or from an hcp to a nearest hcp site. The format of the database is exactly the same as in the fcc-only pattern recognition scheme\cite{slkmc1, pslkmc}, except that the new scheme requires more rings to identify the same neighborhood than does the fcc-only method.

In order to minimize the size of the database we exploit the sixfold symmetry of the fcc(111) surface. In particular, the following five symmetry operations were used in recognizing equivalent configurations: ($1$) $120^\circ$ rotation, ($2$) $240^\circ$ rotation, ($3$) mirror reflection, ($4$) mirror reflection followed by $120^\circ$ rotation and (5) mirror reflection followed by $240^\circ$ rotation. At each KMC step, only if neither a given configuration nor its symmetric equivalent is found in the database, is a saddle-point search carried out to find the possible processes along with their activation energies for subsequent storage in the database. Note that for an atom on an hcp site, the first ring is always equal to $21$ while for an atom on an fcc site it is $42$.

The difference in the value of the first ring for an atom on an fcc and that for hcp site is due to the fact that substrate atoms are oriented differently around the fcc or hcp sites. This fact facilitates in the unique identification of whether an atom is on an fcc or hcp site. Thus for 2-D pattern recognition, inclusion of substrate atoms in the first ring is sufficient. Accordingly all substrate atoms beyond the first ring can be assigned $0$'s instead of $1$'s without any ambiguity in the identification of the local neighborhood. This custom modification to our 2-D simulations reduces computational effort in matching ring numbers when neither fcc nor hcp sites are occupied in a given ring. It also reduces storage of large numbers in the database. For example in the case of a monomer on an fcc site, the ring numbers would be $42$, $0$, $0$ instead of $42$, $273$, $83220$ (as they appear in Fig.~\ref{fig2}(b)).

To find the processes of a given configuration and their respective energy barriers, saddle-point searches are carried out using the drag method. In this method a central atom is dragged (i.e., moved in small steps) towards a probable final position. If the central atom is on an fcc site, then it is dragged towards a vacant fcc site in the second ring; if it is on an hcp site, it is dragged towards a vacant hcp site in the second ring. Since hcp sites are allowed, the atom being dragged from an fcc site to a neighboring one can relax to an hcp site in between the two (Similarly, an atom being dragged from an hcp to a neighboring vacant hcp site can end up on an intermediate fcc site). In other words, processes are possible in which atoms in an island may occupy fcc, hcp or fcc-hcp sites. The dragged atom is always constrained along the reaction coordinate while it is allowed to relax along its other two degrees of freedom (perpendicular to this direction) and all the other atoms in the system are allowed to relax without constraint. Once the transition state is found, the entire system is completely relaxed to find the island's final state. The activation barrier of the process is the difference between the energies of the transition and initial states. We verified the energy barriers of the some of the key processes found by drag method using the nudged-elastic band (NEB) method and found little difference. For inter-atomic interactions, we used interaction potential based on the embedded-atom method (EAM) as developed by Foiles \textit {et al}\cite{Foiles}. In all our calculations we use the same pre-exponential factor ($10^{12}$), this has been proven to be a good assumption for the systems like the present one \cite{handan-prb, handan-ss}.
\begin{figure*}[ht]
\center{\includegraphics [width=15.0cm]{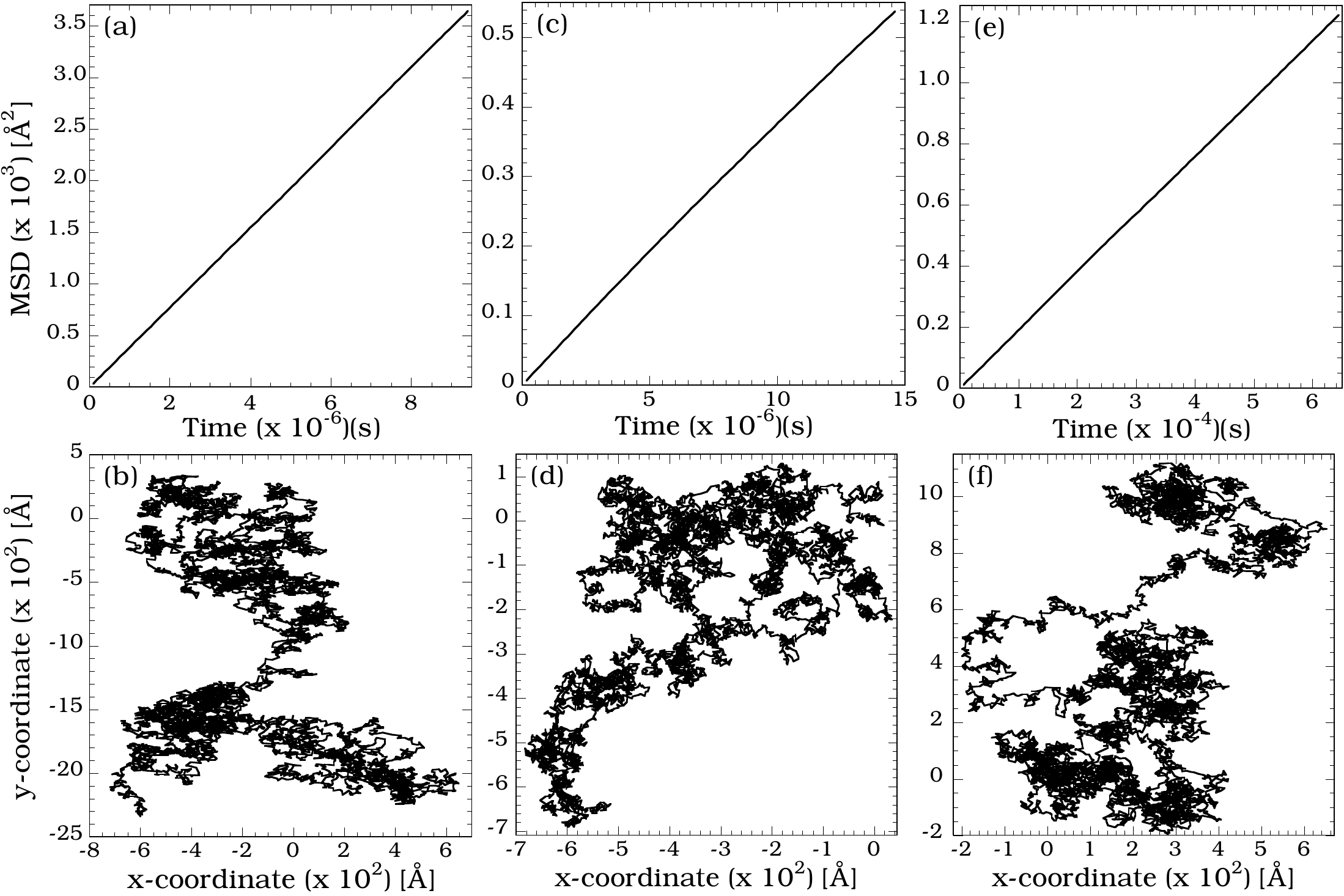} 
\caption{\label{fig3}{Mean square displacements and center of mass trajectories for Cu((a)\&(b)), Ag((c)\&(d)) and Ni(e)\&(f)) at $500$K}}}
\end{figure*}
\begin{table*}[ht]
\caption{\label{table1}Diffusion coefficients at various temperatures and effective energy barriers for Cu, Ag and Ni. Values in brackets are for Cu from reference 10 }

\begin{tabular}{ c | c c c c c c }
\hline
\hline
System &300K & 400K & 500K & 600K& 700K & $E_{eff}$ (eV)\\
\hline
Cu & $1.43\times 10^{6}$~ & $4.53\times 10^{6}$ ~& $3.88\times 10^{6}$~& $1.64\times 10^{9}$~ & $4.82\times 10^{9}$ ~& 0.367 \\ 
& ($7.72\times 10^{4}$) & &$(7.20\times 10^{7}$) & &($1.45\times 10^{9}$) & (0.444)\\
Ag & $6.62\times 10^{4}$ ~& $3.36\times 10^{6}$~& $3.67\times 10^{7}$~ & $1.95\times 10^{8}$ ~& $6.43\times 10^{8}$ ~& 0.414\\
Ni & $5.48\times 10^{2}$ ~& $8.77\times 10^{4}$~& $1.89\times 10^{6}$~ & $1.35\times 10^{7}$ ~& $6.23\times 10^{7}$ ~& 0.525\\
\hline
\hline
\end{tabular}
\end{table*}
\section{Results}\label{results}
We have used SLKMC-II to study the self-diffusion of a 9-atom island of Cu, of Ag and of Ni on the corresponding (111) surface. To account for all types of processes (for compact and non-compact 9-atom islands), especially concerted processes and multi-atom processes, we used $10$ rings to cover the same neighborhood that required $5$ rings in the predecessor scheme, which was capable of recognizing only fcc sites. We begin our simulation for a given system with an empty database. The database is then filled up as new processes are encountered during the simulation. In most cases database accumulation is nearly complete after the first couple of hundred KMC steps, after which the simulation only occasionally performs a calculation of new processes (namely, when a previously unencountered configuration is detected) and stores the result in the database.
In what follows we first present the diffusivities for these three systems at various temperatures and the corresponding effective energy barriers. In discussing the Cu system, we compare our results with those obtained earlier using the fcc-only pattern recognition scheme \cite{slkmc2}. We then highlight some frequently picked processes (and their energy barriers) encountered during these simulations that were impossible to detect under the previous-pattern recognition scheme \cite{slkmc1}.
\subsection{Diffusion coefficients and effective energy barriers}
\begin{figure}[ht]
\center{\includegraphics [width=8.5cm]{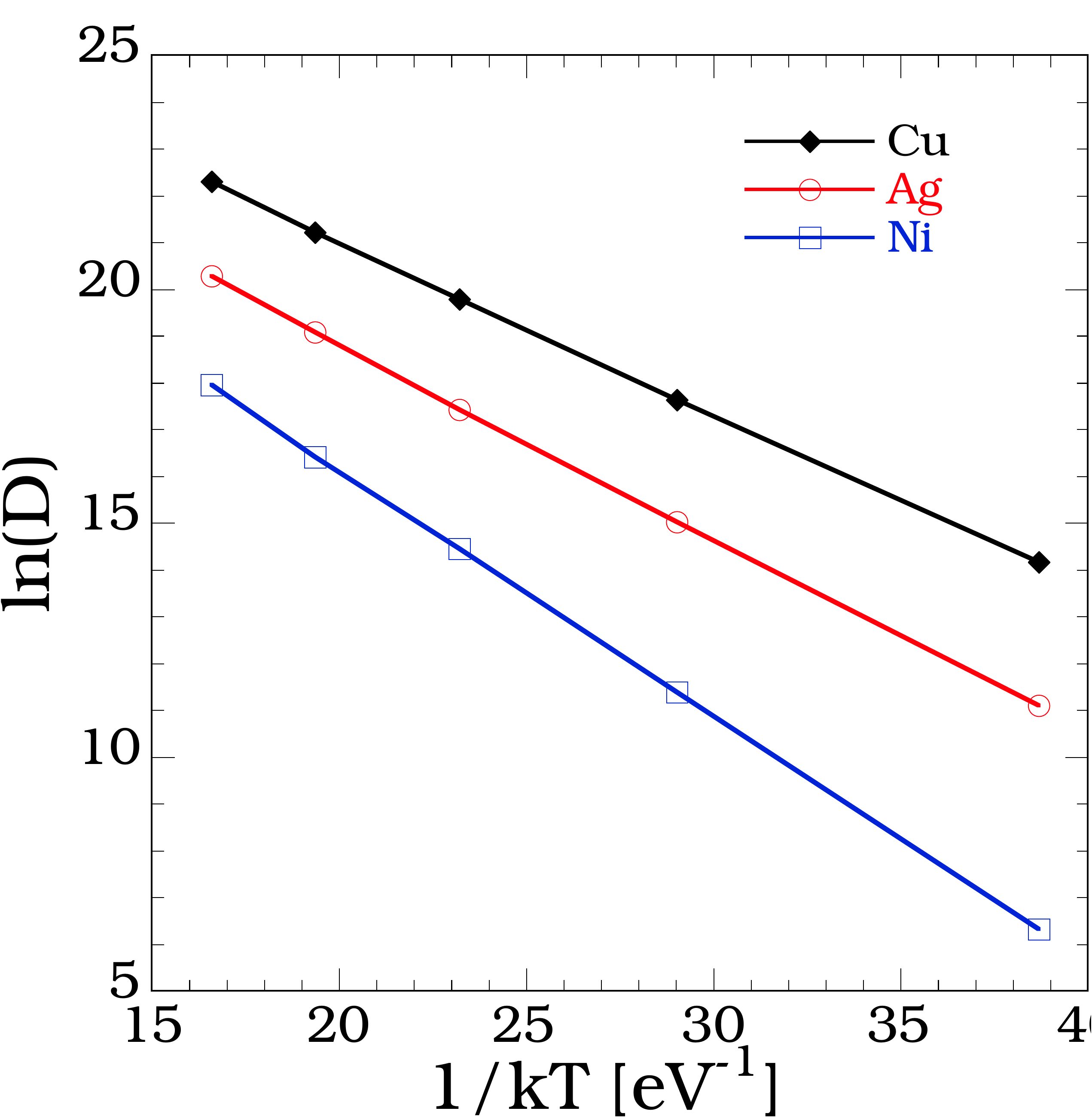} 
\caption{\label{fig4}{Arrhenius plot of diffusion coefficients as a function of inverse temperature for 9-atom Cu, Ag and Ni island. }}}
\end{figure}

We performed SLKMC-II simulations for $10^7$ KMC steps in the case of Cu and Ag and $10^8$ KMC steps in the case of Ni and recorded the trace of the center of mass of a 9-atom island for each system along with the real time of simulations for the temperature range of 300 K-700 K. Fig.~\ref{fig3}(a-f)) shows the mean square displacement and center of mass trajectories for all three systems at $500$ K. Table.~\ref{table1} shows the diffusion coefficients and effective energy barriers derived, respectively, from the mean square displacements (Fig.~\ref{fig3}(a, c $\&$ e)) and Arrhenius plots (Fig.~\ref{fig4}) for each of the three systems. (Values in parentheses there are those for Cu from Ref.~\cite{slkmc2}). Table.~\ref{table2} shows the energy barriers for various types of processes which are picked during the simulations. Some of the details of these types of processes are explained later in the section. We note that for all three systems, a 9-atom island incorporates a compact 7-atom hexagon with two extra atoms at different positions on the boundary of the hexagon (see Fig.~\ref{fig8}). The most frequently-picked processes are the single-atom processes (non-diffusive) -- namely edge running and corner rounding -- whereas kink detachment (diffusive) processes, which are responsible for change in island shape and hence change in the island's center of mass, are less frequent than the most diffusive concerted processes (cf. Fig.~\ref{fig9} $\&$ Fig.~\ref{fig5}). We note that in our simulations of island diffusion, we disallowed any type of detachment processes resulting in the formation of a monomer although they are detected and stored in the database.

As to Cu, from Table.~\ref{table2}, it can be seen that the energy barriers for concerted processes (cf. Fig.~\ref{fig5}(a \& b)) are roughly in the range of $0.265-0.400$ eV (see text below for details), whereas for single-atom edge-running processes (cf. Fig.~\ref{fig9}), the energy barriers are roughly in the range of $0.245-0.285$ eV. Single atom processes, especially edge-running and corner-rounding processes are the most frequently picked processes during simulations. These processes do not contribute to island diffusion except when preceded by kink-detachment processes. The next most frequently picked processes are the concerted processes in which all atoms in an island move together either from fcc to hcp or vice versa. Concerted process cause maximum displacement of the center of mass and hence contribute the most to the diffusion of the island. After concerted processes, multi-atom processes like reptation (cf. Fig.~\ref{fig6}(a \& b)) and shearing (cf. Fig.~\ref{fig7}(a \& b)) contribute most to the island diffusion. These processes are picked only when island becomes non-compact, as happens but rarely in the temperature range under study here. As mentioned earlier, kink detachments which occur very rarely at low temperature becomes the frequent events at high temperatures and are responsible for island diffusion through island shape change. For Cu 9-atom island diffusion we obtained effective energy barrier of $0.370$ eV, which is close to diffusion barriers for concerted processes. Accordingly we conclude that Cu 9-atom island on fcc (111) surface diffuses via concerted processes. As can be seen from Table.~\ref{table1}, because of the inclusion of fcc and hcp adsorption sites in our simulations, diffusivity of Cu is higher and correspondingly effective energy barrier is almost $0.077$ eV less than that obtained using SLKMC\cite{slkmc2} with fcc-only pattern recognition scheme.

For the case of Ag 9-atom island, it can be seen from the Table.~\ref{table2} that the gap in the energy barrier between edge running and concerted processes is large as compared to the case of Cu 9-atom island. Accordingly concerted processes being picked are less frequent for the same number of KMC steps as for Cu 9-atom island resulting in lower diffusivity of Ag 9-atom island (compare Figs.~\ref{fig3} (b)\&(d)). Also kink detachment processes are picked more often as compared to the case of Cu. For Ag 9-atom island diffusion we obtained an effective energy barrier of $0.414$ eV which is close to the energy barrier of concerted processes indicating that island diffusion mainly proceeds via concerted processes.
\begin{table}
\caption{\label{table2} Energy barriers for different types of processes. A(B)2$\rightarrow$A(B)2 represents A(B) edge running, A(B)2$\rightarrow$C1 represents A(B) corner rounding and K3$\rightarrow$C1 represents kink detachment processes}
\begin{tabular}{ c | c c c }
\hline
\hline
Processes &Ni~ (eV)& Cu~(eV) & Ag~(eV) \\
\hline
Concerted & 0.530~ & 0.265-0.400&0.420 \\ 
Reptation & 0.400 ~& 0.320 &0.348\\
A(B)2$\rightarrow$A(B)2 & 0.322(0.440) ~& 0.245(0.285)&0.260(0.330)\\
A(B)2$\rightarrow$C1 & 0.400(0.540) ~& 0.310(0.390)& 0.300(0.350)\\
K3$\rightarrow$C1 & 0.730 ~& 0.590& 0.555\\
\hline
\hline
\end{tabular}
\end{table}

It can be seen from Table.~\ref{table2} that Ni has even larger gap in the energy barrier between edge running and concerted diffusion processes ($0.322$ eV and $0.530$ eV). As a result more number of KMC steps ($10^8$) are needed for the case of Ni 9-atom island diffusion to get good statistics. Comparing Figs.~\ref{fig3} (b), (d) \& (f) it can be seen that diffusivities for Ni is less that Cu and Ag, hence higher effective energy barrier shown in Table.~\ref{table1}. Most frequent compact shape for Ni 9-atom island is 7-atom hexagon with a dimer attached at different locations around it (see Fig.~\ref{fig8} (a \& b)). Whenever this dimer is on the corner of the hexagon, energy barrier for this dimer to go around the corner (dimer shearing) of the compact hexagon is $0.320$ eV and is favored over single atom corner rounding. Although for Ni, kink detachment processes never occur at 300 K, but occurrence of them starts to increase from $400$ K to $700$ K. Since most of the displacement is coming from the concerted motion of the island, effective energy barrier of $0.533$ eV for Ni 9-atom island is close to the energy barrier of concerted diffusion processes.
\subsection{Example Processes}
As mentioned earlier, the database generated by SLKMC-II records all of the single-atom and multi-atom processes captured by its fcc-only predecessor, but register as well a number of processes that were outside that predecessor's grasp: concerted processes, two types of multi-atom processes - shearing and reptation\cite{chirita,hamilton} as well as single atom processes involving moves from fcc to hcp, hcp to fcc and hcp to hcp.
\subsubsection{Concerted Processes}
\begin{figure}[ht]
\center{\includegraphics [width=8.5cm]{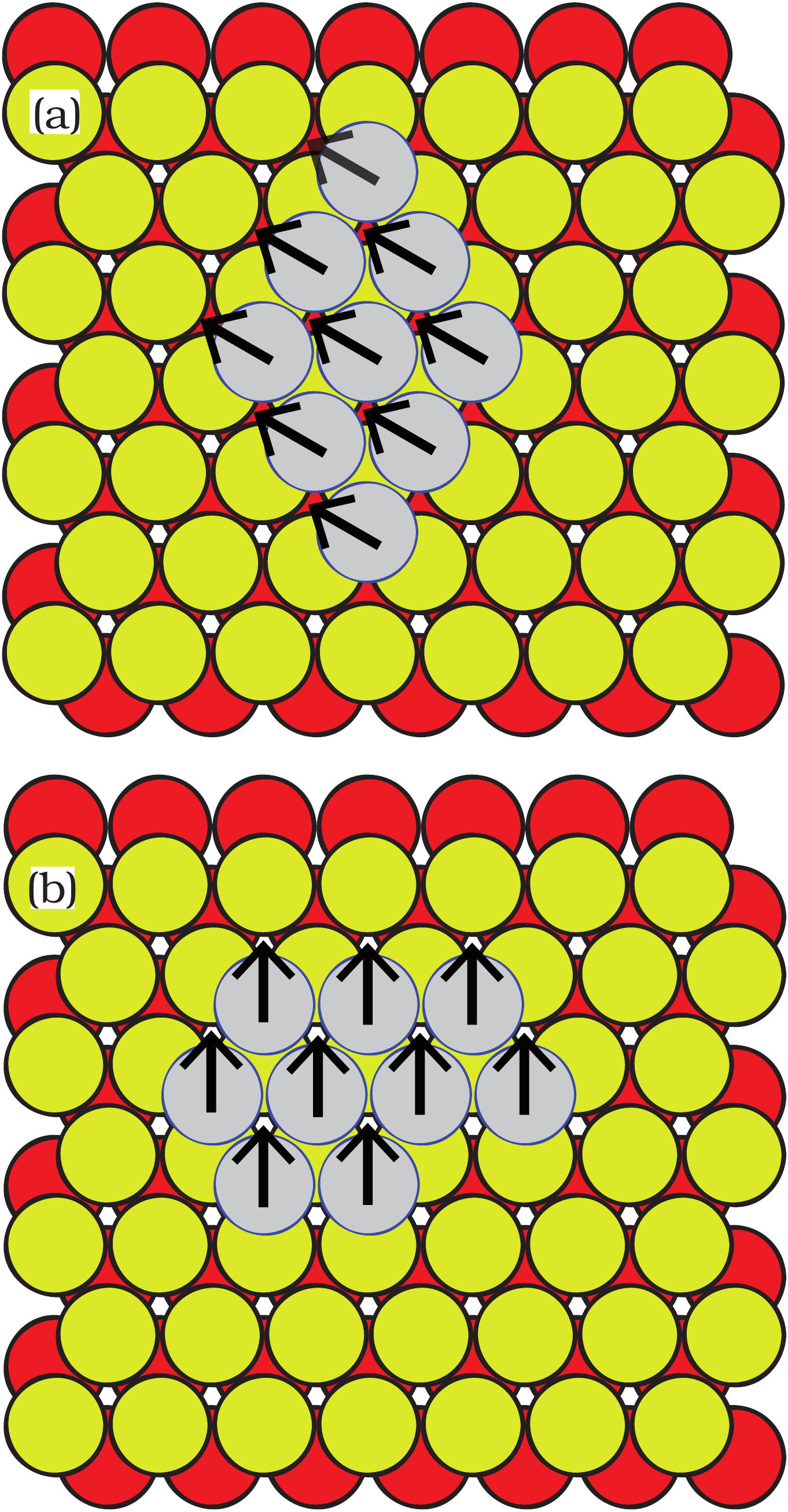} 
\caption{\label{fig5}{Examples of 9-atom islands with compact shape that often undergo concerted processes as marked by the arrows (Only one concerted process is shown for each case). (a) for Cu \& Ag. (b) for Cu \& Ni.}}}
\end{figure}

It is already well known that small islands on fcc(111) surface diffuse via concerted processes,\cite{wang, KK, Li} in which all atoms in the island move simultaneously. In these processes, atoms in the island move from fcc to hcp if the island is initially on fcc or from hcp to fcc if on hcp. SLKMC-II automatically finds all such processes. Interestingly enough all three systems under study here exhibit such concerted diffusion processes. Some examples of compact shapes frequently diffusing through concerted processes for Cu, Ni and Ag are shown in Fig.~\ref{fig5}(a \& b). Energy barriers for these processes are different among the three materials. As can be seen from Table.~\ref{table2}, the energy barrier for concerted processes for the 9-atom Cu island is in the range $0.265 - 0.400$ eV: $0.265$ eV for diffusion from hcp to fcc; above $0.350$ eV for fcc to hcp, depending on the shape of the island; $0.400$ eV for concerted diffusion of non-compact shapes.

For the 9-atom Ag island, the activation barrier for concerted diffusion processes is around $0.420$ eV, with little difference whether the island moves from hcp to fcc or vice versa; non-compact 9-atom Ag island diffuses not through concerted but through multi-atom or single atom processes (discussed in the next sections). The activation barrier for concerted diffusion of the 9-atom Ni island is around $0.530$ eV; as in the case of Ag, it makes little difference whether this diffusion begins on hcp or on fcc sites; and non-compact shapes diffuse not through concerted but through multi- or single-atom processes.
\subsubsection{Multi-atom processes}
\begin{figure}[ht]
\center{\includegraphics [width=8.5cm]{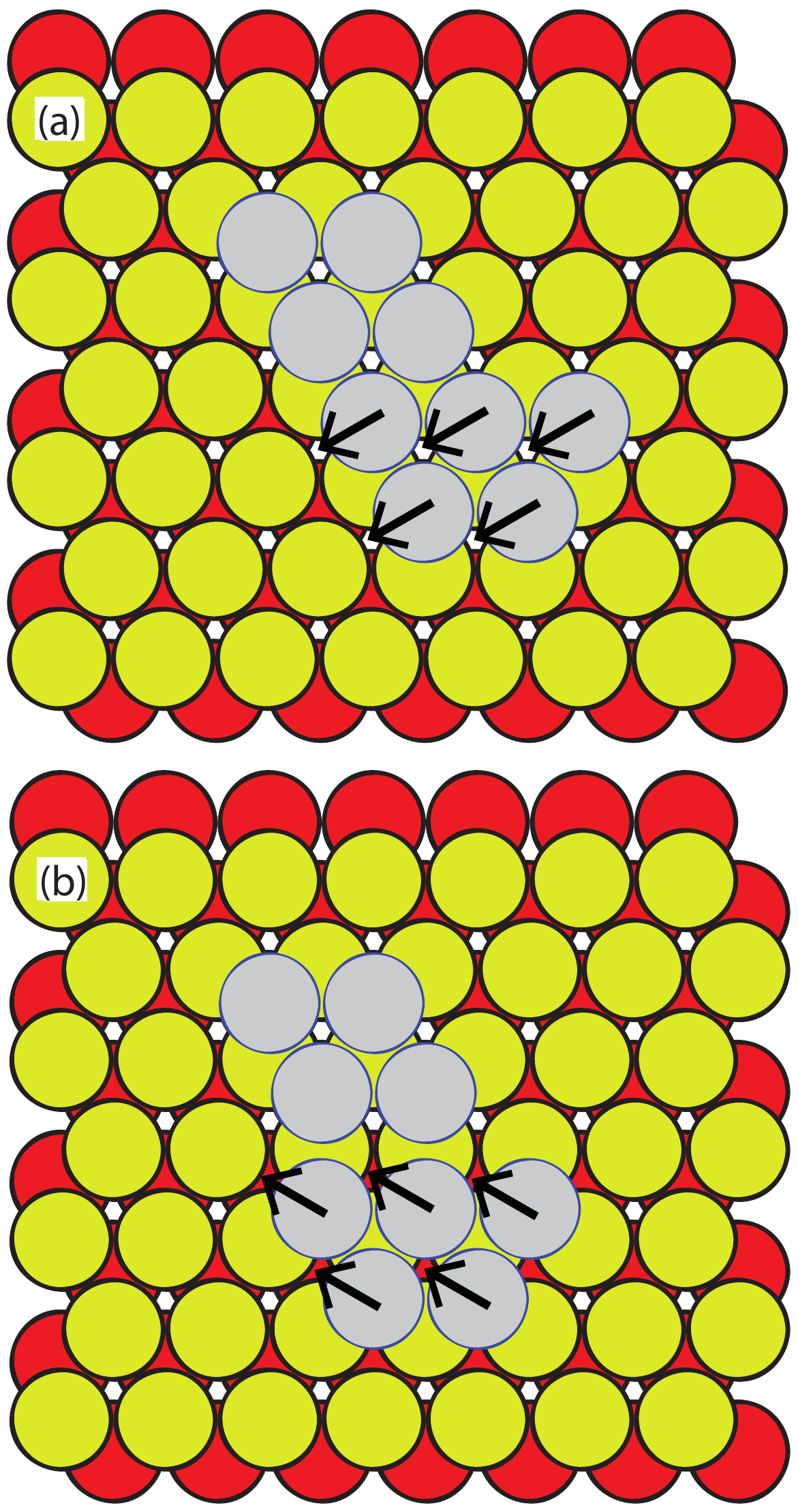} 
\caption{\label{fig6}{An example of reptation mechanism, where part of the island moves from hcp to hcp via fcc in two steps. (a) Step1: Initial state where 9-atom island is on the hcp sites with arrows on 5 atoms showing the reptation direction. (b) Step 2: Final state after step 1 with arrows showing next step of reptation process.}}}
\end{figure}

SLKMC-II finds a variety of multi-atom processes, which can be classified into two types: reptation\cite{chirita,hamilton} and shearing. Shearing is a single-step process: if the island is on fcc sites, some part of it consisting of at least 2 atoms moves to its nearest fcc sites (If the island sits on hcp sites, some part of it larger than a single atom moves to the nearest hcp sites). Reptation, in contrast, occurs in two steps. In the first step part of the island (larger that one atom) moves in such a way as to create a stacking fault (that is, from fcc to hcp or vice versa). In the second step, parts of the island moves in some way that eliminates the stacking fault. Fig.~\ref{fig6} offers an example of a reptation process occurring in a 9-atom island occupying hcp sites. Part of the island, in this case consisting of 5 atoms moves along the path indicated by arrows in Fig.~\ref{fig6}(a) to neighboring fcc sites as shown in Fig.~\ref{fig6}(b). Step two could be completed in four different ways where either the 4 atoms in the island or 5 atoms in the island joins the rest of the island with all atoms occupying the same kind of sites (hcp or fcc). One of the possibilities is shown by arrows in Fig.~\ref{fig6}(b). It should be noted that although energy barrier for reptation processes is lower compared to concerted processes(see Table.~\ref{table2}), these type of processes are possible only if island changes its shape from compact to non-compact, which is rare at room temperature. Contribution of reptation processes to the island diffusion increases with temperature since the probability of island changing into a non-compact shape also increases with temperature. We note that the two steps of reptation process are stored as individual configurations in the database.

\begin{figure}[ht]
\center{\includegraphics [width=8.5cm]{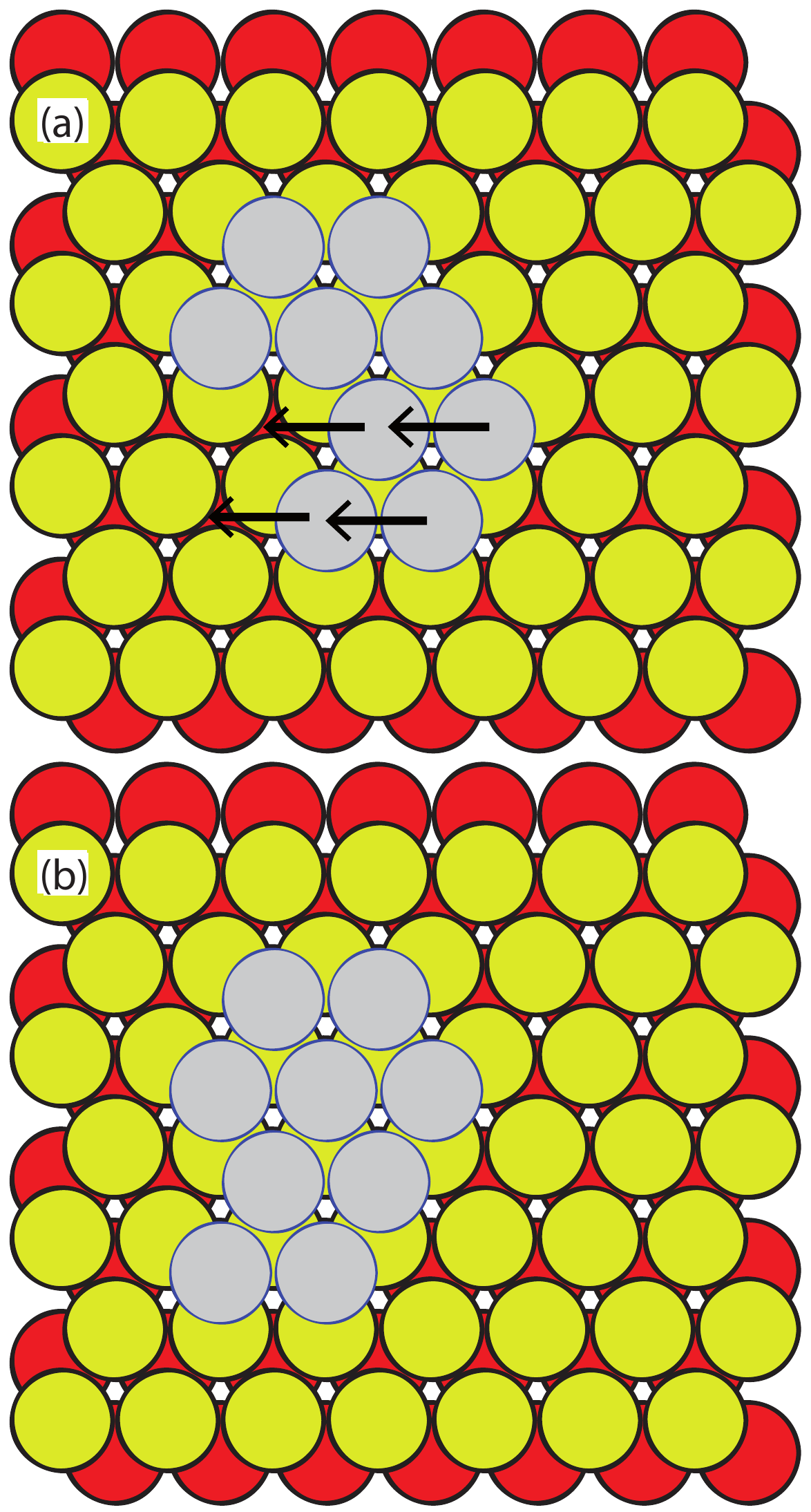} 
\caption{\label{fig7}{Example of a shearing mechanism in which 4 atoms move simultaneously. (a) Initial state of a 9-atom island on hcp sites. The arrows showing the direction of shearing. (b) Final state of the island.}}}
\end{figure}

\begin{figure}[ht]
\center{\includegraphics [width=8.5cm]{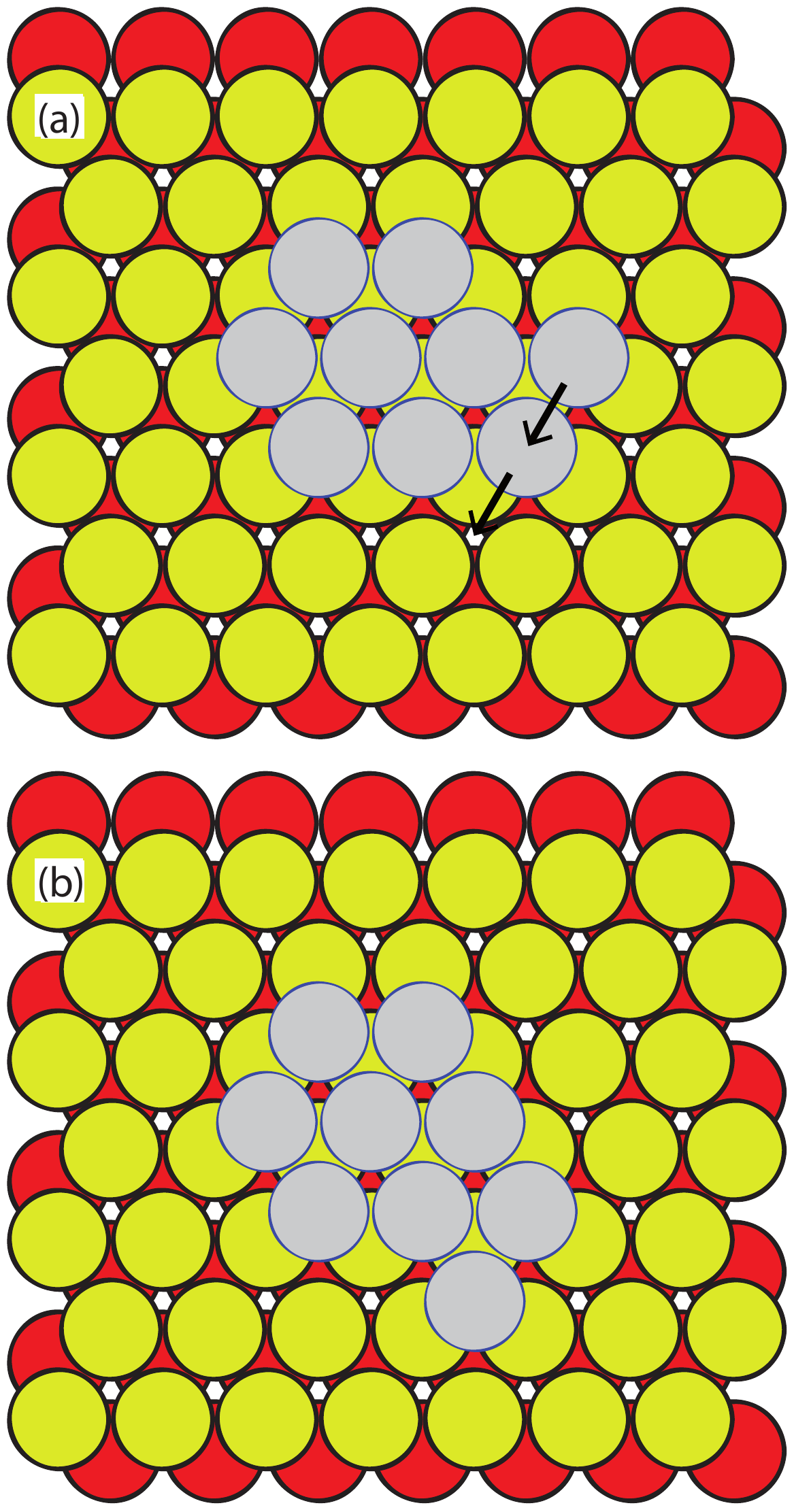} 
\caption{\label{fig8}{ Dimer diffusing around the corner of the 9-atom island. (a) Initial state of 9-atom island on fcc sites with arrows on the lower right dimer on the corner of compact hexagon showing the shearing process. (b) Final state of the 9-atom island after dimer on the corner glides to the other side of the hexagon.}}}
\end{figure}

Fig.~\ref{fig7} shows an example of a shearing process observed during SLKMC-II simulations. In this example, a group of 4 atoms move from hcp to neighboring hcp sites (The simulations also found shearing to occur in islands on fcc sites). A variety of other multi-atom processes are also revealed during our SLKMC-II simulations, specifically, a dimer diffusing around the corner of the island as shown in Fig.~\ref{fig8}. For Ni island diffusion, this process has an activation barriers of $0.320$ eV; for Cu it is $0.460$ eV and for Ag it is 0.470 eV. Although this process is one of the most frequently picked for 9-atom Ni islands, it does not result in diffusion of the island as a whole.
\subsubsection{Single-atom processes}
\begin{figure}[ht]
\center{\includegraphics [width=8.5cm]{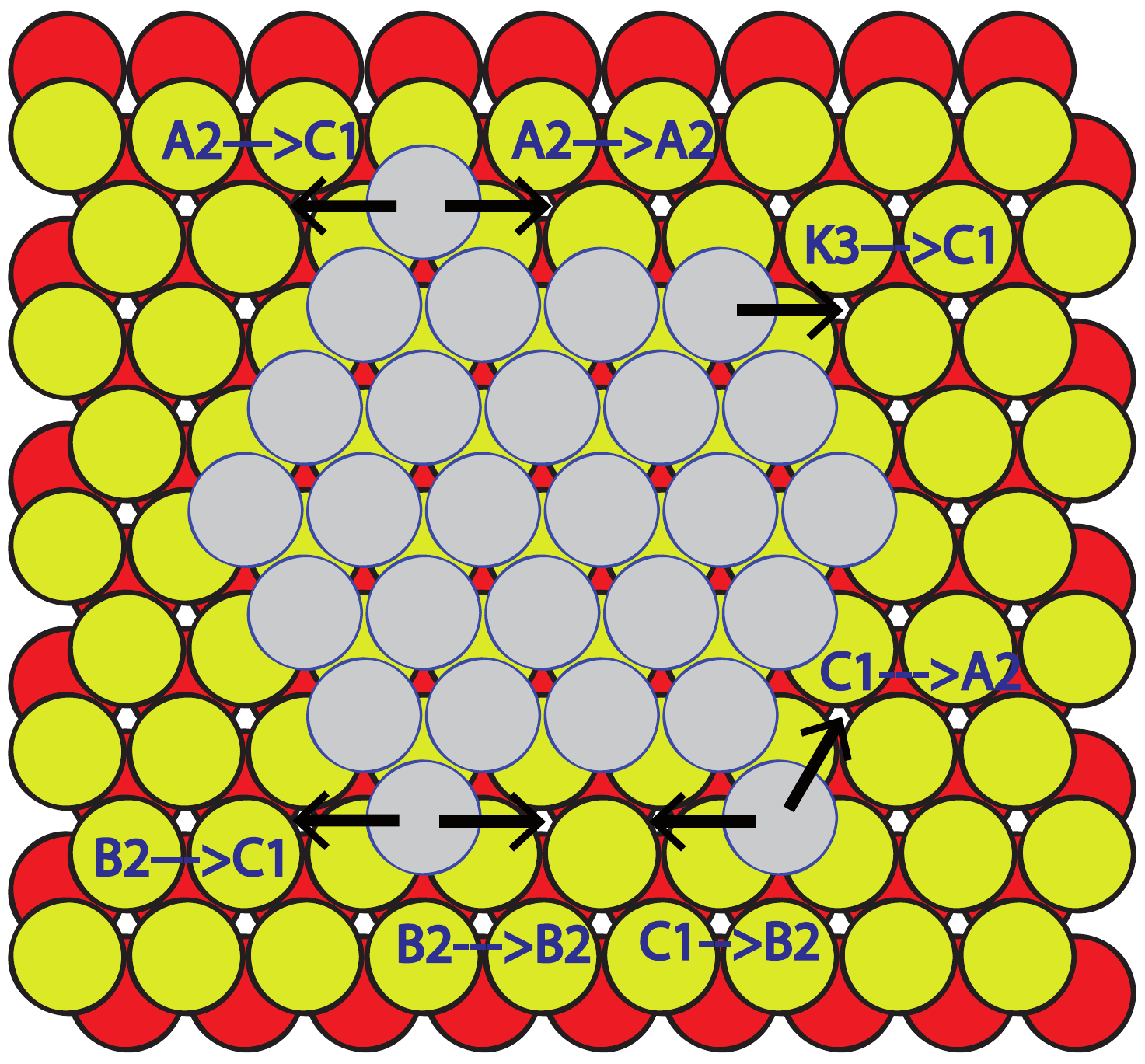} 
\caption{\label{fig9}{Various types of single atom diffusion processes }}}
\end{figure}

All single-atom processes possible for the systems under study -- edge running, monomer detachment from different steps and corners, kink-detachment, kink-attachment, corner rounding and whether fcc to fcc or hcp-hcp were also found and stored during our simulations. On fcc(111) surface, there are two types of microfacets, namely (100) micro-facet (also called A step) and (111) micro-facet (also called B step). In all single atom processes, we use the notation Xn$\rightarrow$Yn, where X and Y can be different types of steps (A, B), Kink (K), Corner (C) and n represents the number of bonds the atom has before and after the process. One type of single-atom process is an atom's movement along an A (alternatively a B) step edge of the island represented in Table.~\ref{table2} as A(B)2$\rightarrow$A(B)2. Cu has the lowest barrier for these 2 types of diffusion processes and Ni the highest, the difference between the barriers along the A- and B-step edge is smallest for Cu (0.040 eV) and largest for Ni (0.118 eV). Another type of single-atom process is corner rounding, denoted in Table.~\ref{table2} as A(B)2$\rightarrow$C1, in which an atom moves from one type of step having 2 bonds to the other (A to B or B to A) by traversing a corner having single bond. Various types of kink attachment and detachment processes, Table.~\ref{table2} includes only one: K3$\rightarrow$C1, in which an atom detaches to form a corner. Our SLKMC-II database incorporates other types of Kink involving processes like C1$\rightarrow$K3, K3$\rightarrow$A(B)2, (A(B)2$\rightarrow$K3) as well. Fig.~\ref{fig9} illustrates various types of single atom processes that find a place in our database.
\section{CONCLUSIONS}\label{SUMMARY AND PROSPECTS}
In KMC simulations, the processes an atom or a collection of atoms can perform depend on the local environment. A pattern-recognition scheme allows KMC simulations to identify this environment, together with the processes possible within it and their respective energy barriers, for on-the-fly storage in and retrieval from a database. Unlike its predecessor, in which only fcc processes were allowed, our new pattern-recognition scheme (which makes possible what we call SLKMC-II) is designed to include processes that involve both fcc and hcp sites on the fcc(111) surface. A key innovation is the inclusion in the recognition scheme of top-layer atoms in the first ring (but only that ring), which enables it to distinguish whether an atom occupies an fcc or an hcp site, and in a way that reduces the bulkiness of the database. There are some trade-offs to be taken into account. Because it requires twice the number of rings to cover the same neighborhood as required by its predecessor, additional computational effort is called for during identification (in matching ring numbers). Still, the new scheme is quite simple and easy to implement – quite apart from its enablement of considerably more realistic simulations.

We have tested this new pattern recognition by studying 2-D self-diffusion of 9-atom islands of Cu, Ag and Ni on fcc (111) surface. These achievements open the way for further development of pattern-recognition strategies in ways that will extend the reach of KMC simulation. 
Although rarely, atoms in small clusters may even in homoepitaxial systems \cite{offkmc} sit on bridge sites. We have found this to be so even in our simulations of small clusters \cite{tobepublished}. One way to incorporate such processes is to resort to an off-lattice pattern-recognition scheme \cite{3Doffkmc}. But a better way would be to further refine the pattern-recognition scheme reported here, so as to include bridge sites as well. Though this approach would increase computational expense because more rings would be required in order to identify the neighborhood than are required for distinguishing between fcc and hcp occupancy, it would still be faster than carrying out off-lattice KMC simulations. Secondly, although the pattern-recognition scheme we have described is essentially a 2-D scheme, we are in the process of extending it to handle on-lattice 3-D processes as well. Thirdly, as a study we are currently completing will show, the pattern-recognition scheme described here is suitable for doing SLKMC simulations of island growth on fcc (111) surfaces. Finally, we note that the scheme we have described for the study of self-diffusion on fcc(111) surfaces can be adapted to the study of other surfaces namely (110) and (100). We can also use this method to study hetero systems where adatom-adatom interactions are weaker than adatom-susbstrate interactions.
\section*{Acknowledgments}
We would like to acknowledge computational resources provided by University of Central Florida. We also thank Lyman Baker for critical reading of the manuscript and Oleg Trushin for helpful discussion at the beginning of this project.  This work was supported by NSF-ITR grant 0840389 and by DOE Grant DE-FG02-07ER46354.
\section*{References}
\bibliography{references}
\end{document}